\documentclass[showpacs,amsmath,amssymb,twocolumn,nofootinbib]{revtex4}
\newcommand{\be}{\begin{equation}}
\newcommand{\ee}{\end{equation}}

\newcommand{\bax}{\bar{x}}
\begin{document}
\title{Vacuum Energy and Standard Model Physics}
\author{Gian Luigi Alberghi}
\affiliation{Dipartimento di Fisica and INFN, Via 
Irnerio 46,40126 Bologna,Italy}
\author{Alexander Y. Kamenshchik}
\affiliation{Dipartimento di Fisica and INFN, Via 
Irnerio 46,40126 Bologna,Italy\\
L.D. Landau Institute for Theoretical Physics of the 
Russian Academy of Sciences, Kosygin str. 2, 
119334 Moscow, Russia}
\author{Alessandro Tronconi}
\affiliation{Dipartimento di Fisica and INFN, Via 
Irnerio 46,40126 Bologna,Italy}
\author{Gian Paolo Vacca}
\affiliation{Dipartimento di Fisica and INFN, Via 
Irnerio 46,40126 Bologna,Italy}
\author{Giovanni Venturi }
\affiliation{Dipartimento di Fisica and INFN, Via 
Irnerio 46,40126 Bologna,Italy}
\begin{abstract}
The conditions for the cancellation of one loop contributions to vacuum energy
(both U.V. divergent and finite) coming from the Standard Model (SM)
fields are examined. It is proven that this is not possible unless one
introduces besides several bosons, at least one massive fermion having mass
within specific ranges. On examining one of the simplest SM extensions
satisfying the constraints one finds that the mass range of the lightest
massive boson is compatible with the present Higgs mass bounds. We do not
consider effects associated with condensates or renormalization group running.
\end{abstract}

\maketitle 
\section{Introduction}
Almost sixty years ago Pauli~\cite{Pauli,Giulini} suggested that the vacuum (zero-point)
energies of all existing fermions and bosons compensate each other. This possibility is
based on the fact that the vacuum energy of fermions has a negative sign
whereas that of bosons has a positive one. 
We note that such an idea is realized in a highly constrained way in
supersymmetric models, although supersymmetry breaking must be
present at probed energies in order to explain observed data.
Subsequently in a series of papers
Zeldovich~\cite{zeld} connected the vacuum energy to the cosmological
constant, however rather than eliminating the divergences through a
boson-fermion cancellation he suggested a Pauli-Villars regularization of all
divergences introducing a spectrum of massive regulator fields.
Covariant regularization of all contributions then leads to finite values for both the energy
density and (negative) pressure corresponding to a cosmological constant.

Rather than use a regularization approach we shall assume that the actual
particle content of a theory is such that U.V. divergences do not appear
insofar bosons and fermions should compensate each other as Pauli suggested.
Indeed we have previously examined~\cite{we} the problem of U.V. divergences
of the vacuum energy for both Minkowski and de Sitter space-times and
formulated the conditions for the cancellation of all divergences.
These conditions lead to strong restrictions on the spectra of possible
elementary particle models. In this note we shall apply 
such considerations to the observed particles of the SM and also study the
finite part of the vacuum energy and the possibility of a cancellation for
this contribution also, so as to obtain a result compatible with the observed value
of the cosmological constant (almost zero with respect to SM particle masses).
The cancellation of all one-loop contributions to the cosmological and
Newton constants was also considered in the context of the induced
gravity approach \cite{Fursaev}. We shall instead consider Einstein gravity and obtain
all constraints in Minkowski space which is implicit if all contributions
(divergent and finite) compensate between fermions and bosons.
Let us illustrate them in order.

The requirement that quartic divergences cancel is just that the numbers of
bosonic and fermionic degrees of freedom be equal ($N_B=N_F$).
The conditions for the cancellation of quadratic and logarithmic divercences
on a flat Minkowski background are 
\begin{equation}
\sum m_s^2 +3 \sum m_V^2 = 2 \sum m_F^2
\label{quad}
\end{equation}
and
\begin{equation}
\sum m_s^4 +3 \sum m_V^4 = 2 \sum m_F^4\,,
\label{log}
\end{equation}
respectively.
Here the subscripts $s$, $V$ and $F$ denote scalar, massive vector and massive
spinor Majorana fields respectively (for Dirac fields it is enough to put $4$
instead of $2$ on the right-hand sides of Eqs. (\ref{quad}) and (\ref{log})). 
For the case of a de Sitter spacetime, equations giving conditions for the
cancellation of quadratic and logarithmic ultraviolet 
divergences are more involved.
Some  examples of these conditions for simple particle physics models have
been presented in Ref.~\cite{we}.  

The requirement that the finite part of the vacuum energy (and pressure) also 
be very small compared with SM masses suggests that we also need a
compensation between the finite parts of fermion and boson vacuum energies, obtaining
\begin{equation}
\sum m_s^4\ln m_s +3 \sum m_V^4 \ln m_V - 2 \sum m_F^4\ln m_F =0.
\label{en-finite} 
\end{equation}
This leads to a zero cosmological constant (Minkowski space).

As is known the observed number of fermionic degrees of freedom in the Standard
Model is much higher than the number of bosonic degrees of freedom \cite{particles}.
Indeed $N_F$ is equal to $96$ (if we consider the neutrinos as massive particles)
while the number of bosonic degrees of freedom, carried by the photon, 
the gluons and the $W^{\pm}$ and $Z^0$ bosons is equal to $27$.
Thus we need an additional 69 boson degrees of freedom, one of which 
is the Higgs boson.
Ideally we would like to obtain some minimal extension of the Standard Model, which
would not modify the fermionic degrees of freedom while just adding hypothetical bosons.

Our main result is a proof that, within the given framework, such an extension does not exist.
In other words, we show that on introducing new bosonic fields, which provide 
the cancellation of the ultraviolet divergences in the vacuum energy density,
the finite part of the effective cosmological constant is always positive
and of order of the mass of the top quark to the fourth power.
This leads to the necessity of introducing new heavy fermions.
Indeed we shall find explicit realizations with zero finite energy once one introduces
at least one fermion with a suitable mass.

Following a general analysis,  for the sake of simplicity we shall consider an
explicit minimal extension of the SM with a few massive bosons and weakly coupled,
practically massless, others so as to satisfy the requirement $N_B=N_F$.
Such a possibility is viable in effective action approaches and, for example, has been
considered recently in such scenarios as unparticle physics~\cite{georgi}.
In this minimal framework we shall analyze the boson masses allowed by the
cancellation constraints.

It is obvious that one may study vacuum energy in the more modern and general
setting~\cite{Visser} of effective actions, renormalization group flows, and even attempt to
include the effects of condensates, however we feel that it is worthwhile to
first examine the full consequences for SM physics of Pauli's original
suggestion.

In next section we study the consequences of our equations for the SM particle
spectrum and in the last section our results are summerized and discussed.

\section{The Standard Model and vacuum energy balance}
Let us begin by observing that the mass of the top quark $m_t \approx 170$GeV
is much higher than the masses of all other fermions
(the bottom quark has the mass
$m_b \approx 4.5$GeV while the mass of the heaviest 
$\tau$-lepton is $m_{\tau} \approx 2$GeV).
Thus, on considering Eqs. (\ref{quad}), (\ref{log}) and (\ref{en-finite}) 
we can limit ourselves to only taking into account the
contributions of the top quark, whose mass is conveniently used 
as the reference unit mass, and of the massive vector bosons.
Then the mass of $W^{\pm}$ bosons is $m_W \approx 0.47 m_t$ while that
of the $Z^0$ boson is $m_Z \approx 0.53 m_t$, with $m_t=1$.    

Quantities describing the contributions of the heavy fermion and boson degrees of freedom 
in the conditions (\ref{quad}), (\ref{log}) and (\ref{en-finite}) are then:
\begin{equation}
R^2 \equiv 12 m_t^4 - 6 m_W^4 -3 m_Z^4 \approx 11.5, 
\label{R-define}
\end{equation}
\begin{equation}
h \equiv 12 m_t^2 - 6m_W^2 - 3 m_Z^4 \approx 9.83,
\label{h-define}
\end{equation}
\begin{equation}
L \equiv 12 m_t^4 \ln m_t^2 - 6 m_W^4 \ln m_W^2 -3 m_Z^4\ln m_Z^2 \approx 0.743.
\label{L-define}
\end{equation}

If we denote the masses squared of some hypothetic massive boson fields by $x_1, x_2,
\ldots, x_n$ ($x_i> 0, \forall i$)
then their values should satisfy the conditions
\begin{equation}
\sum_{i=1}^{n} x_i^2 = R^2, \quad \sum_{i=1}^{n} x_i = h,
\quad \sum_{i=1}^{n} x_i^2 \ln x_i \equiv \phi= L\,.
\label{constraint}
\end{equation}
\noindent
We shall now proceed as follows:
 
- Firstly we shall find a lower bound to
the number of massive boson degrees of freedom due to the first two
constraints in (\ref{constraint}), which define a surface {\cal S} in the space
of the $x_i$;

- Secondly we shall study on {\cal S}, for positive $x_i$, the extrema of the
  function $\phi$ by using the method of Lagrange multipliers;

- Finally we shall obtain for $\phi$, which is tipically larger than L on {\cal S},
  its minimum (and maximum) value as a function of the SM
  particle content plus possible additional fermions in order to investigate
  the conditions for the satisfaction of the last constraint in (\ref{constraint}).  

The first two conditions in (\ref{constraint})  have a simple geometrical
sense~\cite{we}: they describe a sphere and a plane in the $n$-dimensional
space and their intersection {\cal S} is an $(n-2)$- dimensional
sphere, eventually to be sliced on the positivity boundary of the $x_i$. 
The distance of the plane from the origin of the
coordinates is $h/\sqrt{n}$. In order to have an intersection between
the sphere of radius $R$ and the plane it is then necessary to have
\begin{equation}
n > \frac{h^2}{R^2} \approx 8.4.
\label{condition5}
\end{equation}
Thus, the number of massive bosonic degrees of freedom should at least be
equal to $9$. In general it is convenient to introduce the integer
value $n_0$ for such a threshold
\begin{equation}
n_0 = \left\lfloor \frac{h^2}{R^2} + 1 \right\rfloor,
\label{n0}
\end{equation}
so that $n\ge n_0$ is the requirement to have a non empty {\cal S}.
Now, in order to see when the last eq. in (\ref{constraint}) is also satisfied,
it is convenient to calculate the minimum value  of the function
$\phi=\sum_{i=1}^{n} x_i^2 \ln x_i$ on the constraint surface {\cal S}.
 Let us consider an auxiliary function
\begin{equation}
F(\{x_i\}) =  \sum_{i=1}^{n} x_i^2 \ln x_i -
\lambda \left(\sum_{i=1}^{n} x_i^2 - R^2\right) - \mu \left(\sum_{i=1}^{n} x_i - h\right),
\label{Lagrange}
\end{equation}
where $\lambda$ and $\mu$ are the Lagrange multipliers. 
Searching for the extrema of the function $F$ implies we should equate its
derivatives with respect to $x_i, \lambda$ and $\mu$ to zero. The
last two conditions $\partial F/\partial \lambda$ and $\partial F/
\partial \mu$ again give the first two constraints in (\ref{constraint}).
Differentiation with respect to $x_i$ gives the system of equations:
\begin{equation}
x_i^2 \ln x_i -x_i - 2\lambda x_i - \mu = 0,\ \
i=1,\cdots,n .\label{system}
\end{equation}
Without loss of generality we can choose $x_1 \neq x_2$. Indeed it is
possible to have $x_1 = \cdots = x_n$ if and only if $h^2/n^2 =
R^2$, but this is a degenerate case, when the sphere and plane
touch each other in only one point.

On substituting the values of $x_1$ and $x_2$ into the first two
equations of the system (\ref{system}), one obtains
$\bar{\lambda}$ and $\bar{\mu}$ as functions of $x_1$ and
$x_2$:
\begin{equation}
\bar{\lambda} = 1 + \frac{x_1\ln x_1^2 - x_2\ln x_2^2}{x_1 - x_2},\quad
\bar{\mu} = \frac{x_1 x_2 (\ln x_2^2 - \ln x_1^2)}{x_1 - x_2}.
\label{lambdamu}
\end{equation}

Let us suppose that $\bax_1,\cdots,\bax_n,\bar{\lambda},\bar{\mu}$ are a
solution of the system (\ref{system}) on {\cal S}, i.e.
with the first two Eqs in (\ref{constraint}) already satisfied.
On now substituting these values of $\bar{\lambda}$ and
$\bar{\mu}$ into the $n-2$ remaining equations of the system
(\ref{system}) we can easily see that a solution is given by
$x_1=x_3 = x_4 = \cdots = x_{k+1}$ and $x_2=x_n=x_{n-1}=\cdots=x_{k+2}$.
This solution is a stationary point of the
function $F$, or in other words the conditional stationary point
of the function $\phi$. Such a solution, with $k$ coordinates having the value
$\bax_1=x$ and the remaining $n-k$ coordinates the other value $\bax_2=y$,
is given, as function of $k$ and $n$, by
\begin{equation}
x=x(k,n) = \frac{h}{n} + \sqrt{\frac{R^2(n-k)}{nk} - \frac{h^2(n-k)}{n^2k}}\,,
\label{solx1}
\end{equation}
\begin{equation}
y=y(k,n) = \frac{h}{n} - \sqrt{\frac{R^2 k}{n(n-k)} - \frac{h^2 k}{n^2(n-k)}}\,,
\label{soly1}
\end{equation}
where $1 \leq k \leq n-1$.

The values of $x$ given by Eq. (\ref{solx1}) are always positive,
while the values of $y$ can be negative.
It is easy to show that the condition for the positivity of $y$ is 
\begin{equation}
k < n_0 \leq n \,.
\label{positive}
\end{equation}
We have seen that points of the type described above always
satisfy the stationarity conditions  (\ref{system}) on the constraint surface.
This does not mean that stationary points of other types cannot exist.
Indeed, the  analysis of the structure of Eq. (\ref{system}) shows
(see for details \cite{future}) that, in principle, 
stationary points whose coordinates $x_i$ have {\it three} different
values can exist. However, if such points exist, at least one of these three
values is negative and, hence, is of no interest to us. Thus, the minimum of
the function $\phi$ can be reached {\it only} for the stationary points having
the coordinates (\ref{solx1}), (\ref{soly1}) or on the boundary of the
positivity region, where at least one of the coordinates $x_i$ is equal to
zero. For this last case the problem is reduced to one with lower dimensionality $n$.    

If $n=n_0$ (the smallest possible value for the dimensionality
of $n$) we notice that on the surface {\cal S} all the $x_i$ have positive values.
Thus, the maximum and minimum values of the function $\phi$ on the constraint
surface are obtained only for one of the pairs of points with the coordinates $x$
and $y$ (see formulae (\ref{solx1}), (\ref{soly1})). 

Furthermore the following more general statement is true: for $n \ge n_0$
the maximum value of the function $\phi$ corresponds to $n=n_0$ and $k = 1$
while its minimum value corresponds to the point with $n=n_0$ and $k = n_0-1 $.
To prove it one may compute the derivatives of the function  
\begin{equation}
\phi_1(k,n) = k\, x^2\ln x + (n-k)\, y^2 \ln y,
\label{phi1}
\end{equation}
with respect to $k$ and n. It can be shown~\cite{future} that $d\phi_1/dk<0$
and $d\phi_1/dn>0$ for the range of possibile physical values of $k$ and $n$. 
In particular this means that the function $\phi_1(k,n)$ decreases with
increasing $k$ and has its minimum value at $k = n_0-1$ and $n=n_0$.
This minimum value is 
\begin{eqnarray}
\phi_{1\, min}=\bar{\phi}_1(R,h) = \!\!\!\!\!&&(n_0 -1)x^2(n_0-1,n_0)\ln
x(n_0-1,n_0) \nonumber \\
&& + y^2(n_0-1,n_0)\ln y(n_0-1,n_0)
\label{minim}
\end{eqnarray}
where one has to use Eqs. (\ref{n0}), (\ref{solx1}) and (\ref{soly1}).
The solution of the equation 
\begin{equation}
\sum_{i=1}^{n} x_i^2 \ln x_i = L 
\label{eqbase}
\end{equation}
exists on the constraint surface {\cal S}  only if 
\begin{equation}
\bar{\phi}_1(R,h) < L.
\label{maincondition}
\end{equation}
A direct calculation shows that for $n_0 = 9, \phi_{1\min} \approx 1.95$ which
is higher than $L \approx 0.743 $ in Eq. (\ref{L-define}). For $n>n_0$
the minimum value of the function $\phi_{1}$ is higher, so our first result is
that it is not possible to have the cancellations in an extension of the SM
obtained on only introducing new bosonic fields.

Let us now consider an extension which includes new fermionic fields with masses $m_f$ in units
of top mass and $n_f$ degrees of freedom. On introducing the new parameters
\begin{eqnarray}
&&\tilde{R}^2=R^2+\sum_f n_f \,m_f^4\nonumber \\
&&\tilde{h}=h+\sum_fn_f \,m_f^2\nonumber \\
&&\tilde{L}=L+\sum_fn_f \,m_f^4\,\ln m_f\,,
\label{ext_ferm}
\end{eqnarray}
one may search for solutions allowed by the inequality
$\phi_1(k=n_0-1,n)< \tilde{L}<\phi_1(k=1,n)$ for
different $n \ge n_0$, by varying $n_f$ and $m_f$. In
general the lower bound for $n$ slightly increases from $n_0$
depending on $n_f$ and $m_f$. 
In the simplest case with only one additional fermion we already find solutions.
Let us here give the results for
three interesting cases and a non physical ($n_f\to\infty$) case:
\begin{eqnarray}
&&{\rm Majorana}: \quad n_f=2, \quad m_f>1.52 \nonumber \\
&&{\rm Dirac}: \quad n_f=4, \quad m_f>1.46 \nonumber \\
&&{\rm Dirac \, quark}\, (3\, {\rm colors}): \,n_f=12,\, m_f \in
[0.3536,0.3592]\cup\nonumber \\
&&\,\,\,\,[0.4,0.655]\cup[0.6892,0.6914]\cup[1.4, \infty[ \nonumber \\
&& n_f \to \infty\,,\quad m_f \in [\frac{1}{\sqrt{n_f}},0.853]\cup[1.355,\infty[
\label{ext_ferm_masses}
\end{eqnarray}
More possibilities are allowed if one introduces more fermionic fields
with different masses.

We finally investigated a simple extension of the SM obtained on introducing a Majorana
fermion and assuming for the bosonic sector $n\ll N_B$, so that most of
the bosons are practically massless and weakly coupled, as in the unparticle
scenario mentioned near the end of the previous section.
We found, as a function of the Majorana fermion mass, sets of
solutions for the massive bosons (we consider the case $n=10$).
Of particular interest is the lightest boson mass which
could play the role of the Higgs mass. In such a case we find for it the
allowed mass intervals (in GeV): $[111,139]$, $[115,172]$, $[112,178]$ and
$[86,177]$ for $m_f^2=2.5,\, 3,\, 3.5$ and $4$ $m_t^2$
respectively.
These are solutions compatible with the actual Higgs mass limits~\cite{particles}.

\section{Conclusion}
We have proved that it is impossible to construct a minimal extension of the
SM by finding a set of boson fields which, besides cancelling the ultraviolet
divergencies, can compensate the residual huge contribution of the known
fermionic and bosonic fields of the Standard Model to the finite part of the
vacuum energy density.

On the other hand we have found that the addition of at least one massive
fermionic field is sufficient for the existence of a suitable set of boson fields
which would permit the cancellations and we have obtained the allowed windows
for the masses.
This result is by itself very suggestive since in extensions of the SM often new
extra fermions are considered, independently of any cancellation requirement.
An example is the explanation by a see-saw mechanism of the smallness of the
neutrino masses, which requires the presence of high mass
Majorana neutrinos.

The addition of one Majorana or Dirac fermion requires a mass roughly at
least $50$\% higher than the top quark mass.
If the fermion belongs to a new quark family, new mass windows
appear (see (\ref{ext_ferm_masses})) with a range of values lower than the
top quark mass, implying that this low mass family should be weakly coupled. 
Furthermore we have investigated numerically the simplest extensions of the SM
which satisfy the constraints and found that the lightest massive boson can
have a mass compatible with the bounds on the Higgs boson mass.

The problem we have addressed could also be studied in the context of
renormalized effective actions wherein all parameters are running, including the
cosmological constant as well as the field masses.
Further the cosmological constant may also be affected by the presence of
condensates which we do not consider in our analysis.
Another point we feel is important, but not understood, is a finite vacuum energy
contribution due to the presence of bound states, originating from interactions, whose
conceptual distinction from fundamental particles may not appear obvious when
one is trying to give an effective formulation at different scales. This
is a typical feature of interacting quantum field theories wherein ``fundamental'' degrees
of freedom appear to be different for different scales, one example being given by
strong interaction physics. Again we believe that the most promising approach
is given by a RG flow of effective actions. 
In this sense our results are only a first step in a more general scheme,
which also takes into account any form of interaction.

Nonetheless our results, which are compatible with the present data,
are encouraging since they suggest that reasonable non supersymmetric
extensions of the SM with almost zero vacuum energy may exist.
 
\noindent
{\bf Acknowledgements }\   
We are grateful to G.E. Volovik for useful correspondence. A.Y.K. is thankful
to A.M. Akhmeteli for fruitful discussions.

\end{document}